\documentclass[superscriptaddress,twocolumn,10pt,nofootinbib,floatfix]{revtex4-1}
\usepackage{amssymb}
\usepackage{amsmath}
\usepackage{appendix}
\usepackage{times}
\usepackage{graphicx}
\usepackage{subeqnarray}
\usepackage{bbold}
\usepackage{enumerate}

\usepackage{color}
\usepackage{bm}

\begin{document}
\title{Orbital shift-induced boundary obstructed topological materials with a large energy gap}

\author{Ning Mao}
\author{Runhan Li}
\author{Ying Dai}
\email{daiy60@sdu.edu.cn}
\author{Baibiao Huang}
\affiliation
{School of Physics, State Key Laboratory of Crystal Materials, Shandong University, Jinan 250100, China}
\author{Binghai Yan}
\affiliation
{Department of Condensed Matter Physics, Weizmann Institute of Science, Rehovot, Israel}
\author{Chengwang Niu}
\email{c.niu@sdu.edu.cn}
\affiliation
{School of Physics, State Key Laboratory of Crystal Materials, Shandong University, Jinan 250100, China}

\begin{abstract}
We propose boundary obstructed topological phases caused by Wannier orbital shift between ordinary atomic sites, which, however, cannot be indicated by symmetry eigenvalues at high symmetry momenta (symmetry indicators) in bulk. 
On the open boundary, Wannier charge centers can shift to different atoms from those in bulk, leading to in-gap surface states, higher-order hinge states or corner states. To demonstrate such orbital-shift-induced boundary obstructed topological insulators, we predict eight material candidates, all of which were overlooked in present topological databases.
Metallic surface states, hinge states, or corner states cover the large bulk energy gap (for example, more than 1 eV in TlGaTe$_2$) at related boundary, which are ready for experimental detection. Additionally, we find these materials are also fragile topological insulators with hourglass like surface states.
\end{abstract}

\maketitle
\date{\today}

\noindent{\bf KEYWORDS:} second-order topology, third-order topology, higher-order topology, surface states, hinge states, corner states, boundary obstruction \\

While boundary conditions play rich implications for both fundamental physics and potential applications, the boundary obstruction is currently introduced for the topological classification of quantum matters, where even if two Hamiltonians are adiabatically equivalent under periodic boundary conditions, they cannot be connected for certain surface terminations under open boundary conditions~\cite{boundary,superboundary,ezawaboundary,khalaf2021boundary}. Different from the conventional topological insulators~\cite{hasan22,qi2011topological,Bansil2016rmp,Ando2015,Xiao2021NRP}, the boundary obstructed topological insulators (BOTIs) are not captured by bulk energy gap closing in periodic boundary conditions, revealing a hidden topology of overlooked trivial insulators~\cite{tqc1,Po2017,Song2018,Watanabe2018,Tang2019,nature1,nature2,nature3}. Indeed, topology of conventional trivial insulators currently matures into a significant burgeoning field, describing an extension and fine-graining of the topological paradigm~\cite{xu2021three,xu2021filling,Nelson2021PRL}. For instance, obstructed atomic insulators (OAIs) and orbital-selected OAIs (OOAIs) form as results of the spatial obstruction and representation obstruction, respectively~\cite{schindler2021non}. And remarkably, they have emerged as promising candidates for many interesting properties including low work function, strong hydrogen affinity, and electrocatalysis~\cite{unconventional,oaicatalytic}. Moreover, the boundary-obstructed topological high-temperature  superconductivity is proposed in iron pnictides, opening a new arena for highly stable Majorana modes~\cite{boundary,superboundary}.

Characterized with generalized bulk-boundary correspondence, on the other hand, higher-order topological insulators (TIs) have expanded the topological classification and recently drawn significant attentions~\cite{Benalcazar61,highorderfirst,frankbi,higherordereuln2as2, highorderbieuo,Yue2019,Park216803,frankbi,kempkes2019robust,highordergranpj,highordergraprl,highordersplit}. For which, the $n$th-order TIs in $d$ dimensions host protected features, such as hinge or corner states, at $(d-n)$-dimensional boundaries. Currently, both the second- and third-order TIs have been experimentally confirmed in a variety of metamaterials~\cite{Garcia2018,Imhof2018,Peterson2018,xue2019acoustic,ni2019observation,noguchi2021evidence}. However, for electronic materials, realizations of higher-order TIs are limited to second-order ones, and the third-order TI is still missing. Therefore, a general method towards predicting higher-order TIs with realistic material candidates is highly desirable. Interestingly, systems having boundary obstructions are characterized either by hinge or corner states, arising as a result of quantized electric quadrupole or octupole moments~\cite{multipolescience,benalcazar2017electric,khalaf2021boundary}. Such electric multipole moments origin from the Wannier orbitals shift between an occupied Wyckoff position and empty Wyckoff position for the BOTIs. Therefore, a natural question arises as to whether the Wannier orbitals shift between occupied Wyckoff positions, previously regarded trivial phase, can result in the emergence of hidden topology and even with either hinge or corner states.

\begin{figure}
\centering
\includegraphics{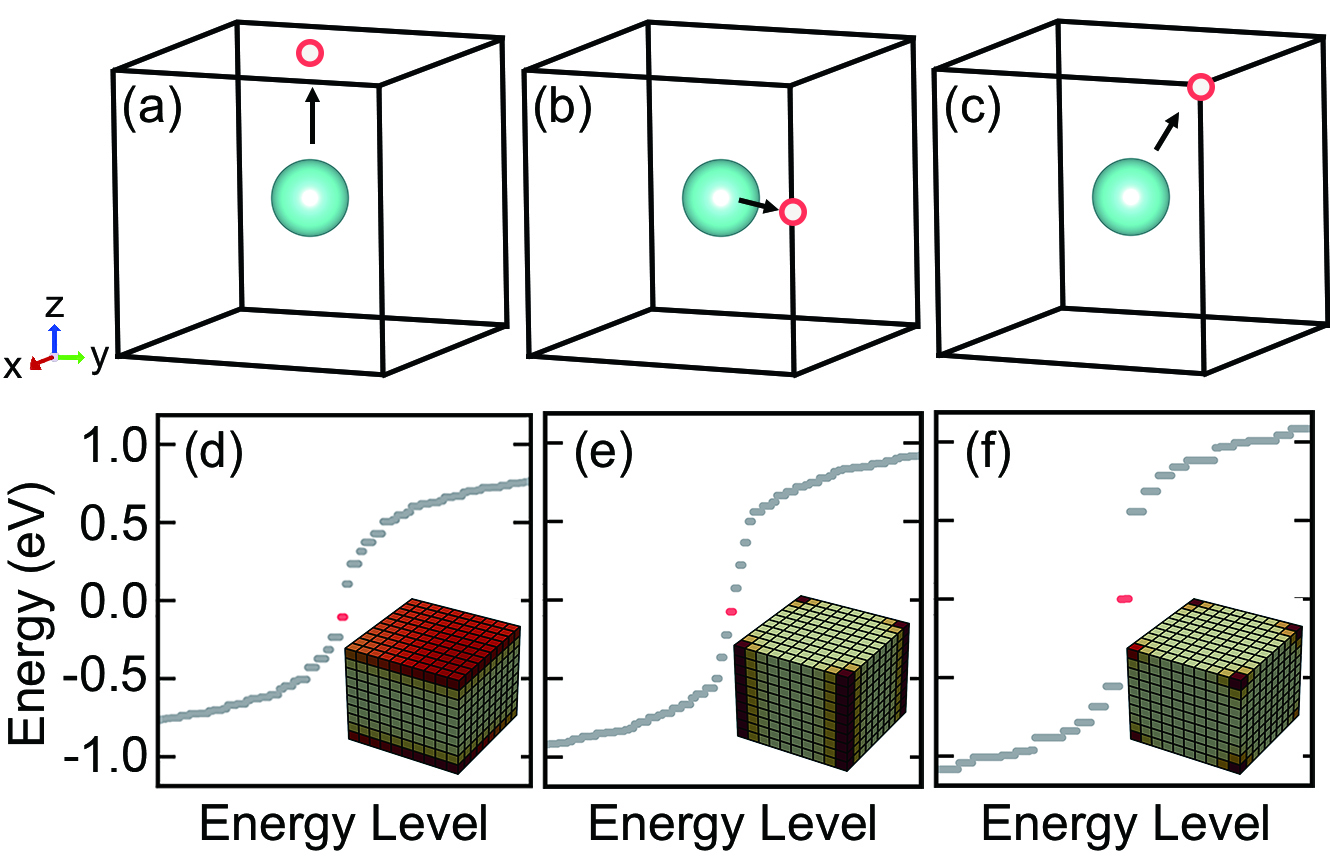}
\caption{ Sketch of the boundary states versus the boundary polarization. The unit cell contains one cyan atom in the center of cubic lattice, and Wannier orbital of atom shift from cyan atoms to red circle along the direction of black arrow. When the Wannier orbital shift into (a) plane, (b) hinge, and (c) corner, the (d) surface states, (e) hinge states, and (f) corner states emerge, respectively.}
\label{high_model}
\end{figure}

In present work, we put forward the realization of orbital shift-induced BOTIs with fragile topology, namely boundary obstructed fragile TIs (BOFTIs). As a consequence of boundary obstruction, metallic surface states and higher-order hinge or corner states can emerge regardless of spin-orbit coupling (SOC) or bulk band inversion. Under periodic boundary condition, BOFTIs exhibit trivial symmetry indicators and  trivial real space invariants~\cite{tqc1,tqc3}. Under certain open boundary condition, at least one of Wannier orbitals will deviate from the original Wyckoff position and shift into another occupied Wyckoff position, giving rise to induced surface states. Based on the topological quantum chemistry (TQC) theory which is actually beyond symmetry indicators~\cite{tqc1}, we provide a general principle to generate the higher-order states, which depend on the boundary polarization direction, 
by comparing the occupied outer-shell Wannier orbitals in compounds with free atomic orbitals (see Figure.~\ref{high_model}). Furthermore, we identify eight experimentally feasible material candidates of BOFTIs by using the first-principles calculations in space group $I4/mcm$. Besides the metallic surface states, five of them possess symmetry protected hinge states while the other three exhibit nontrivial corner states, rendering them the second- and third-order TIs, respectively.  

\begin{figure}
\centering
\includegraphics{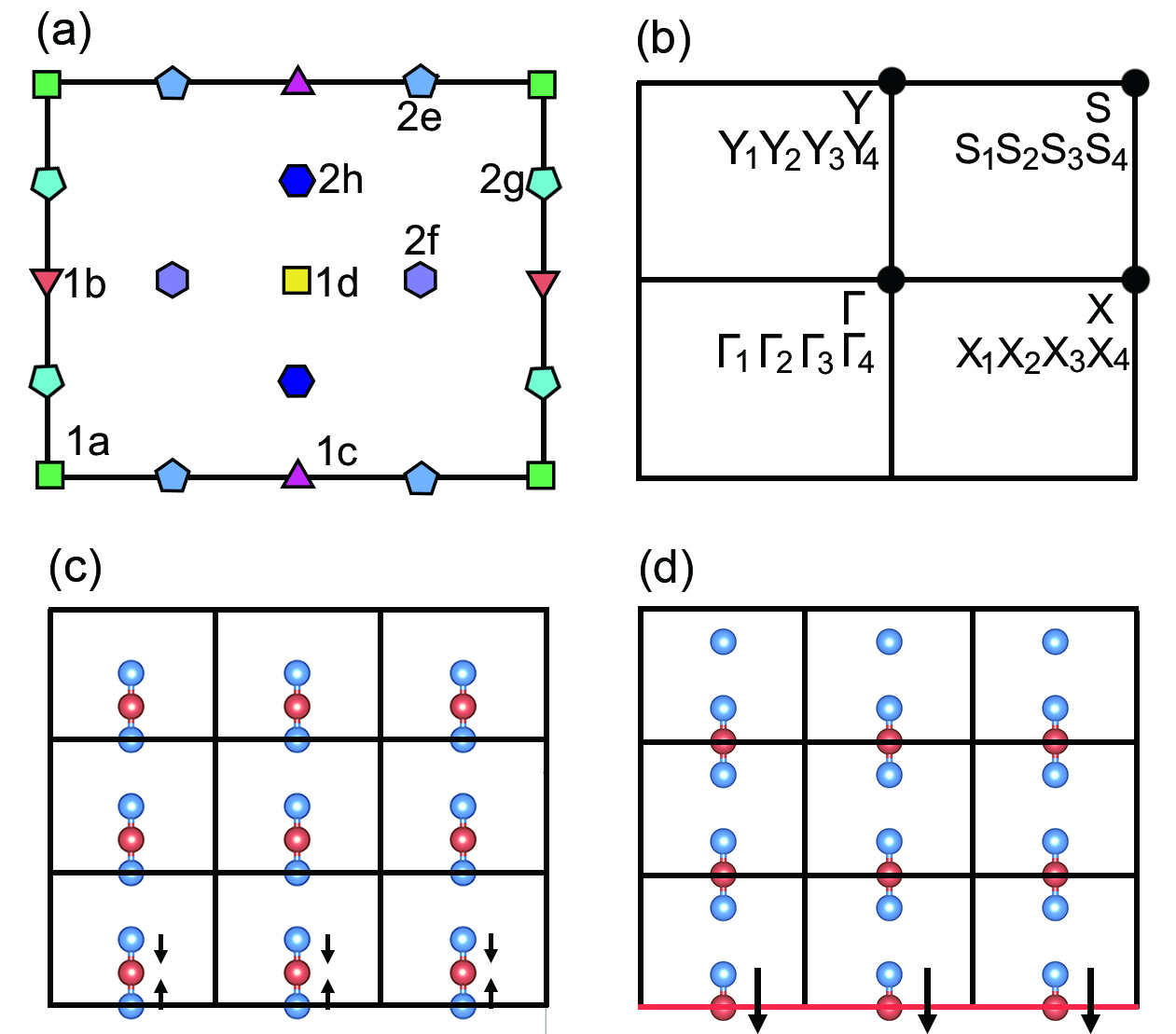}
\caption{(a) Wyckoff positions of the space group $pmm2$. (b) Brillouin zone with the decomposition of Bloch bands into irreducible representations. Real-space configurations terminated for (c) blue and (d) red atoms.}
\label{model}
\end{figure}

We demonstrate the orbital-shift-induced BOTI with a cubic space group $Pmm2$ (No.25) or equivalently the layer group $pmm2$ (No.23), which is generated by two symmetry operations, $\mathcal{C}_{2z} (x\rightarrow-x, \ y\rightarrow-y)$ and $ \mathcal{M}_y (y\rightarrow-y)$. There are four inequivalent $C_{2z}$-invariant maximal Wyckoff positions, 1a, 1b, 1c, and 1d as illustrated in Figure.~\ref{model}(a). Besides, there are four $\mathcal{M}_{x/y}$-invariant non-maximal Wyckoff positions, 2e, 2f, 2g, and 2h. For an atomic insulator, both the Bloch and Wannier states are trapped in the Wyckoff positions~\cite{cano2021topology}. The atomic orbital $\mu$ on a Wyckoff position j is denoted as $|\boldsymbol{R}j\mu \rangle$ with $\boldsymbol{R}$ being the position of unit cell, and the Wannier state of $|\boldsymbol{R}j\mu \rangle$ is constructed by the Fourier transformation,
$\left|W_{\boldsymbol{R}, \alpha}\right\rangle=\sum_{\boldsymbol{R}^{\prime} j \mu} S_{j \mu, \alpha}\left(\boldsymbol{R}-\boldsymbol{R}^{\prime}\right)\left|\boldsymbol{R}^{\prime} j \mu\right\rangle.$ In the following, we will show that boundary obstruction can also give rise to surface and higher-order states, forming into a novel quantum state of BOFTI. Under periodic boundary condition, BOFTI have the same symmetry characters as atomic insulators who have localized Wannier charge centers (WCCs) at the corresponding atom. Therefore, BOFTI cannot be identified by the symmetry indicators or real space invariants~\cite{song2020twisted,po2017symmetry,Song2018}. However, on the open boundary condition, the WCCs of BOFTI will shift into another orbitals that does not belong to the original orbital, giving rise to a nonzero boundary polarization. When the boundary polarization is vertical to the plane, metallic surface states may show up in this plane. While, if the boundary polarization is vertical to the hinge or corner of the structure, symmetry-protected hinge states or corner states manifest as a result of electric quadrupole or octupole moment (see Figure.~\ref{high_model} and the model analysis in Supplemental Material~\cite{sm}).

Here, we consider three atoms, which are located at two Wyckoff positions as 1d (1/2, 1/2, 1/2) and 2h (1/2, $\pm$y, 1/2). Without considering the effects of SOC, $\mathbb{T}^2$ = 1, the single-valued irreducible representations (irreps) for 1d are A$_1$, A$_2$, B$_1$, B$_2$ and for 2h are $\rm A'$, $\rm A''$~\cite{sm}. The $\rm A'$ and $\rm A''$ are mutually conjugated under the symmetry $\mathcal{C}_{2z}$, meaning that, if we put an atomic orbital that transforms as irreps $\rm A'$ at (1/2, y, 1/2), denoted as $| 0 h A' \rangle_L$, there must be another atomic orbital located at (1/2, -y, 1/2) transforms as irreps $\rm A''$, denoted as $ | 0 h A'' \rangle_R$. Given the fact that group $m$ is a subgroup of $mm2$, the subduction and induction relations can be deduced as follows:
\begin{align}
\begin{aligned}
A_1 \downarrow m &= A' \\ 
A_2 \downarrow m &= A''\\
B_1 \downarrow m &= A''\\
B_2 \downarrow m &= A'\\
A'  \uparrow mm2 &= A_1 + B_2\\
A'' \uparrow mm2 &= A_2 + B_1.\\
\end{aligned}
\end{align}
Such relations explicitly show us that all four atomic orbitals at 1d, $ | W_{0 d A_1} \rangle$, $ | W_{0 d A_2} \rangle$ , $ | W_{0 d B_1} \rangle$, and $ | W_{0 d B_2} \rangle$, can adiabatically move to 2h and form two atomic orbitals $ | W_{0 h A'} \rangle_L$ and $ | W_{0 h A''} \rangle_R$. For which, any one of the four Wannier functions cannot be reduced to preserve the equivalence of 2h and 1d. Therefore, a real space invariant is defined as~\cite{xu2021filling,song2020twisted,xu2021three} 
\begin{align}
\delta_1 = m(A_1) + m(B_2) - m(A_2) - m(B_1),
\end{align}

\begin{figure*}
\centering
\includegraphics{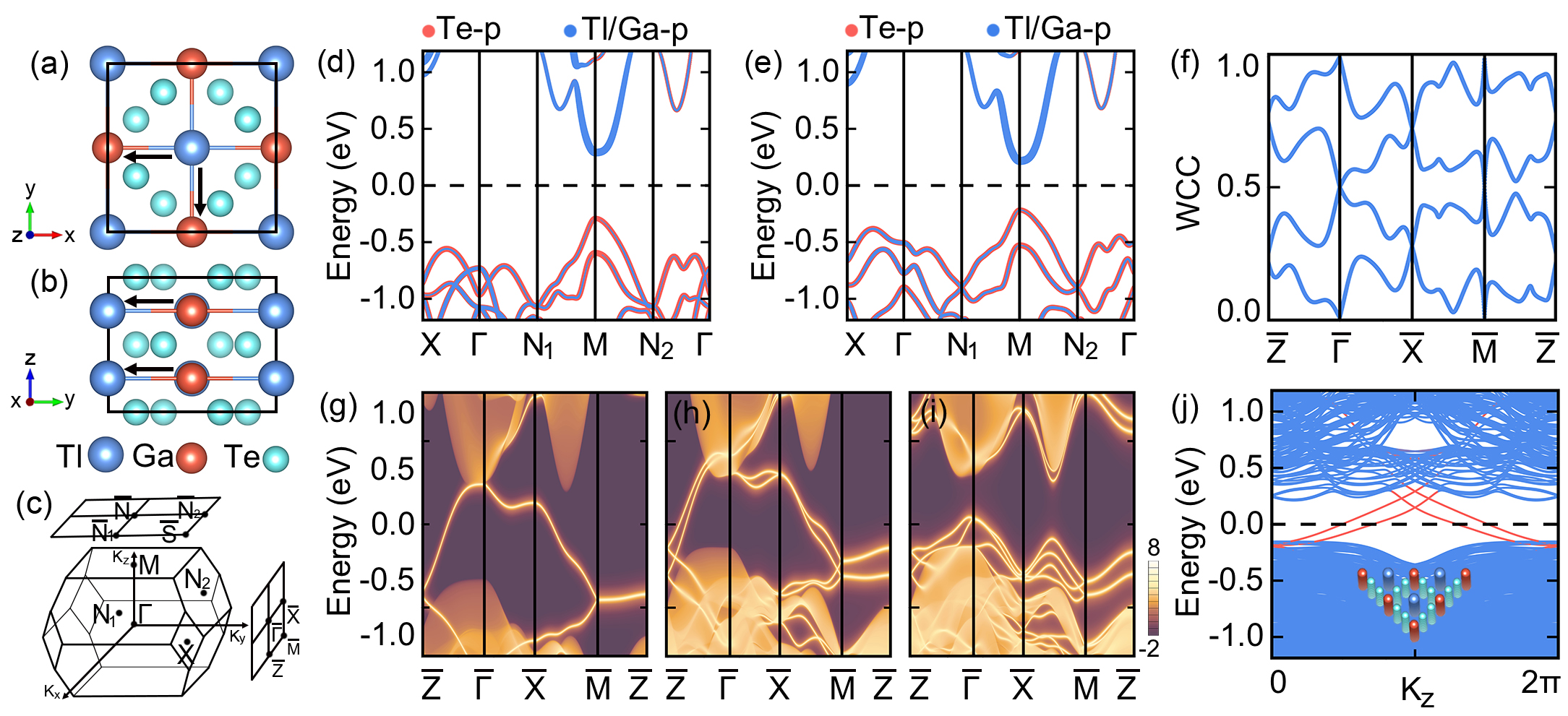}
\caption{(a) Top and (b) side views of TlGaTe$_2$ in space group $I4/mcm$. (c) Brillouin zones for the bulk, (001)-, and (010)-surfaces. Orbitally resolved band structures for TlGaTe$_2$ (d) without and (e) with SOC, weighted with the contribution of Te-$p$ and Tl/Ga-$p$ states. The Fermi level is indicated with a black dashed line. (f) The [010]-directed Wilson loop spectra for the top four isolated valence bands. Metallic surface states of semi-infinite (010)-surface (g) without and (h) with SOC, terminated with the Tl-Ga-Tl atoms. (i) Gaped surface states of  semi-infinite (010)-surface with SOC, terminated with the Te-Te atoms. (j) The band structure of TlGaTe$_2$ nanorod along k$_z$ with the hinge modes indicated in red colors. Inset shows the boundary termination of nanorod.}
\label{second}
\end{figure*}

At first, we put four isolated atomic orbitals $ 2| 0 h A' \rangle$, $ 2| 0 h A'' \rangle$ in the Wyckoff position 2h, and they will spread to form four separated energy bands with elementary band representations (EBRs) $2A'@2h$ and $2A''@2h$ in the momentum space. These four bands correspond to the natural atomic limit with Brillouin zone (BZ) decomposition as illustrated in Figure.~\ref{model}(b), and give rise to a combination of EBRs as $2A'@2h + 2A''@2h = \Gamma_1 + \Gamma_2 + \Gamma_3 + \Gamma_4 + X_1 + X_2 + X_3 + X_4 + Y_1 + Y_2 + Y_3 + Y_4 + S_1 + S_2 + S_3 + S_4$. When atoms are brought together to form a crystal and the Wannier states move to 1d as assumed, the $ | W_{0 d A_1} \rangle_L, | W_{0 d A_2} \rangle_L, | W_{0 d B_1} \rangle_L$ and $ | W_{0 d B_2} \rangle_R$ are formed with a combination of EBRs $A_1@1d + A_2@1d + B_1@1d + B_2@1d = \Gamma_1 + \Gamma_2 + \Gamma_3 + \Gamma_4 + X_1 + X_2 + X_3 + X_4 + Y_1 + Y_2 + Y_3 + Y_4 + S_1 + S_2 + S_3 + S_4$. Clearly, these Wannier states share the same symmetry characters with the isolated ones, and thus leads to the same trivial symmetry indicators and trivial real space invariant $\delta_1 = 0$. Therefore, one can see that BOTIs can be adiabatically equivalent to an atomic insulator in periodic boundary conditions. However, remarkably, metallic boundary states can still be obtained even in the absence of SOC as a consequence of open boundary condition, giving rise to the exotic nontrivial topology.

To show this explicitly, two distinct boundary terminations are considered. While the vertical and horizontal edges lie at $x = 0$ and $y = 0$ that intersect at Wyckoff position 1a, as illustrated in Figure.~\ref{model}(c), the Wannier states can deform from 2h to 1d similar to the periodic bulk state as discussed above. Every $| W_{ \boldsymbol{R} h A'} \rangle_L$ is connected with a $ | W_{\boldsymbol{R} h A''} \rangle_R$ passing through the 1d. Since two positions of 2h are not distinguished, there is no boundary polarization, leading to a trivial termination without any boundary state. On the other hand, without loss of generality, when the edges intersect at 1b with $x = 0$ and $y = 1/2$ shown in Figure.~\ref{model}(d), the opposite horizontal edges are distinctly different owing to the fact that the Wyckoff position 1d can not be shared by two unit cells on opposite edges. That means the red electrons emerge only on one horizontal edge ( $y = 1/2$, marked with red line), but not on the opposite one ( $y = 7/2$). The absence of $ | W_{y=-1 h A'} \rangle_L$ obstruct the Wannier state at 2h to adiabatically evolve to the 1d, leading to an unavoidable Wannier state at 1d. Considering the initial atomic orbital, along the direction tangent to the edge $y = 1/2$, the boundary polarization emerges, which gives rise to the exotic edge states serving as a hallmark of BOFTI. Thus, BOFTIs represent a nontrivial phase with localized Wannier functions, while manifest surface states depending on the boundary terminations. 

Armed with the above definition, we then establish the material realization of BOFTIs with second-order topology in space group $I4/mcm$ (No.140). Figures~\ref{second}(a) and~\ref{second}(b) present the top and side views of ternary thallium chalcogenide, TlGaTe$_2$, which crystallizes in the tetragonal structure with 8 atoms in the primitive unit cell. 2 Tl atoms occupy the Wyckoff position 4a, 2 Ga atoms occupy the 4b, and 4 Te atoms occupy the 8h~\cite{wyckoff, transport}. Each Ga atom is surrounded by four nearest-neighboring Tl atoms within the same layer. The orbitally resolved band structures of TlGaTe$_2$ without and with SOC are illustrated in Figures.~\ref{second}(d) and~\ref{second}(e). The band gaps are 581 and 433 meV, respectively, with the band contribution remains almost the same around the Fermi level before and after inclusion of SOC, which enlarge to 1.17 and 1.03 eV as further checked by the hybrid functional calculations~\cite{hse06}. One important aspect to highlight here is that there is no SOC-induced band inversion, i.e., band gap closing and reopening process, in TlGaTe$_2$, revealing that there is no topological phase transition and thus exhibits the same topological character for TlGaTe$_2$ without and with SOC.

According to TQC theory and WCC calculations~\cite{sm}, the occupied bands of TlGaTe$_2$ are a combination of EBRs, which can be written as $A_1@4a + A_1@4b + B_2@4b + E@4b + A_1@8h + B_1 @ 8h$. Therefore, the occupied outer-shell atomic orbitals for Tl, Ga and Te are $s^2, s^2p^6$, and $s^2p^2$, respectively. Comparing to the $s^2p^1, s^2p^1$, and $s^2p^4$ for Tl, Ga, and Te elements, the electron transfer occurs from Tl-$p$ orbital to Ga-$p$ orbital. Remarkably, such an electron-transfer-process contributes a boundary polarization with both x and y components, implying that TlGaTe$_2$ is potentially a BOFTI with metallic surface states on the (100) and (010) surfaces, and even with the second-order nontrivial topology.

\begin{figure*}
\centering
\includegraphics{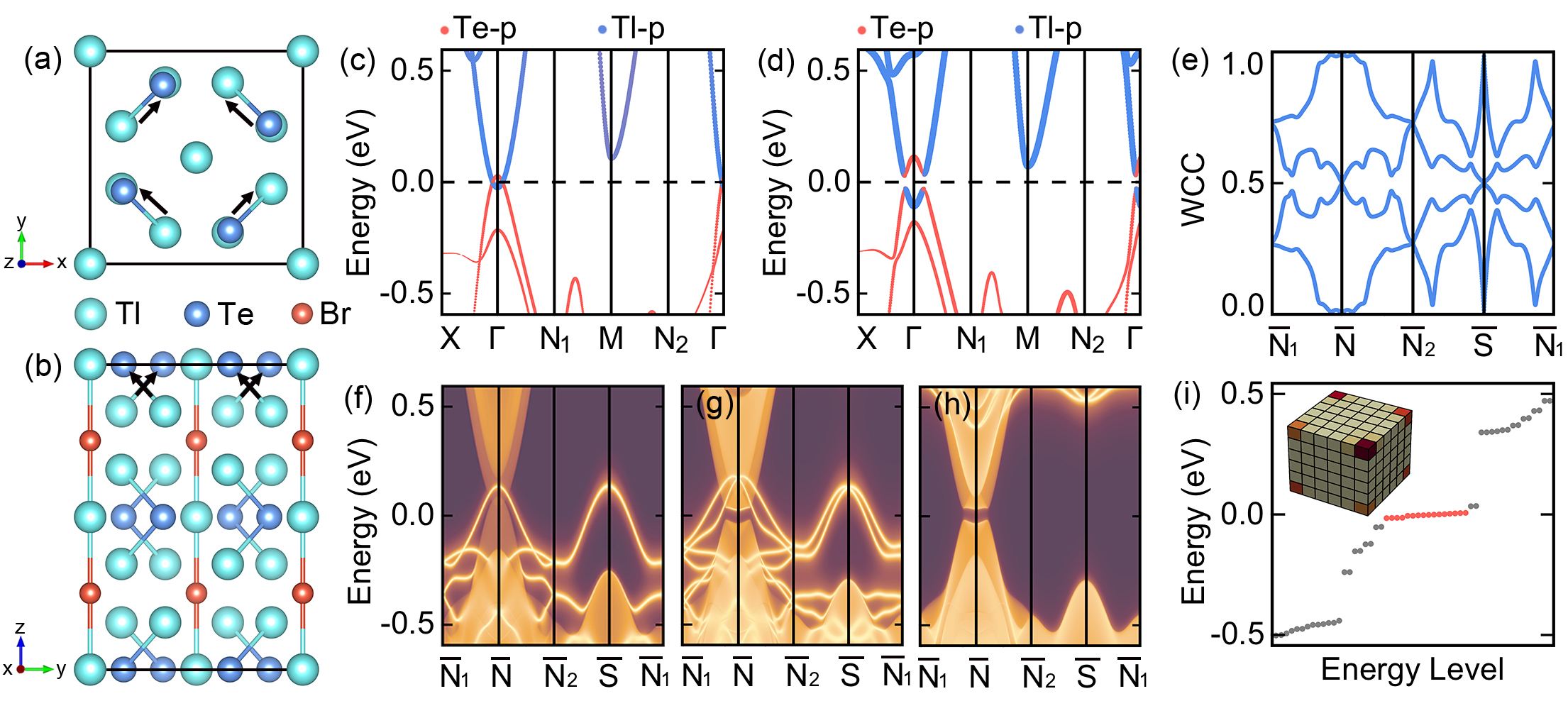}
\caption{(a) Top and (b) side views of Tl$_5$Te$_2$Br in space group $I4/mcm$. Orbitally resolved band structures for TlGaTe$_2$ (c) without and (d) with SOC, weighted with the contribution of Te-$p$ and Tl-$p$ states. The Fermi level is indicated with a black dashed line. (e) The [001]-directed Wilson loop spectra for the top four isolated valence bands. (001)-surface band structure of Tl$_5$Te$_2$Br calculated using surface Green's functions for top surface without (f) or with (g) SOC, terminated with Tl-Te-Te-Tl-Te-Te-Tl atoms. (h) Gaped surface states of (001)-surface for bottom surface with SOC, terminated with Tl-Tl-Tl-Tl atoms. (i) Energy levels of a finite lattice composed of 6 $\times$ 6 $\times$ 6 conventional unit cells for Tl$_5$Te$_2$Br, terminated with Tl-Tl, Tl-Tl, and Tl-Br-Tl-Br-Tl atoms. Inset shows the probability of corner states, where we sum all contributions of probability in one unit cell to represent with a cubic.}
\label{third}
\end{figure*}

To confirm the band topology, we implement the direct calculations of WCCs along $k_x$-, $k_y$-, and $k_z$-directions by the Wilson loop method~\cite{wcc}. Localized groups of Wilson bands are obtained, as shown in Figure.~S6\cite{sm}, indicating trivial symmetry indicators of both the $\mathbb{Z}_2$ and $\mathbb{Z}_8$ invariants. The trivial Wilson bands are consistent with the analysis of EBR, all serving as a signal of atomic insulator for TlGaTe$_2$. Besides, the BR of top four isolated valence bands is a combination of EBRs with positive integer, different from the negative coefficient reported in Refs.~\cite{fragile3,fragile4}. However, as shown in Figure.~\ref{second}(f), the nontrivial winding of Wilson loop along a bent path demonstrates the fragile topological nature of theses four isolated bands~\cite{fragile1,fragile2}.

Remarkably, the metallic surface states of TlGaTe$_2$ always emerge regardless of the SOC, as shown in Figures.~\ref{second}(g) and~\ref{second}(h), although the specific band dispersions are different with respect to $\mathbb{T}$ and glide mirror symmetry $\mathbb{G}_{x} = \{\mathbb{M}_{x}| 0 0 \frac{1}{2}\}$. Along $\mathbb{G}_{x}$ invariant lines $\bar{\Gamma}\bar{Z}$ and $\bar{M}\bar{X}$, one has $\mathbb{G}_{x}^2 =  e^{-ik_{z}}$ in the absence of SOC, and thus the eigenvalues of $\mathbb{G}_{x}$ will be $g_{\pm x} = \pm e^{-ik_z/2}$, namely $\pm 1$ at $\bar{\Gamma}$, $\bar{X}$ and $\pm i$ at $\bar{Z}$, $\bar{M}$. Meanwhile, the $\mathbb{T}$ imposes further constraints, guaranteeing the Kramers degeneracy of $\bar{Z}$ and $\bar{M}$. Moreover, the surface bands along $\bar{M} \bar{Z}$ are always doubly-degenerate due to the protection of $\mathbb{T} \mathbb{G}_{x}$. Consistent with our model analysis, the surface nodal line appears only on one surface but vanishes on the opposite one as a result of the boundary obstruction (see Figure.~S7)~\cite{sm}. After taking SOC into consideration, the glide mirrors satisfy $\mathbb{G}_{x}^2 = - e^{-ik_{z}}$, results in the eigenvalues of $g_{\pm x} = \pm i e^{-ik_z/2}$, i.e., $\pm 1$ at $\bar{Z}$, $\bar{M}$ and $\pm i$ at $\bar{\Gamma}$, $\bar{X}$. In this case, the partner switching of Kramers pairs between $\bar{\Gamma}$ and $\bar{Z}$ ($\bar{M}$ and $\bar{X}$) enforces a band crossing, constituting the so-called hourglass surface states as presented in Figure.~\ref{second}(h). Interestingly, when glide mirror symmetry remains under perturbations, the exotic hourglass-like boundary states are quite robust and survive even they are no longer connecting the conduction and valance bands~\cite{sm}. Moreover, due to the phase factor of glide mirror symmetry, the periodicity becomes $4\pi$, rather than $2\pi$ for a mirror symmetry. The formed surface states are called as M{\"o}bius states as shown in Figure.~S16~\cite{sm}. As presented in Figure.~S8~\cite{sm}, the Dirac crossings near the $\bar{\Gamma} \bar{Z}$ are projected into two intertwined rings. It's noted that the spin is locked to the carrier momentum, rendering the spin-momentum locking effect. While, two non-degenerate rings with the left-handed and right-handed helicities emerge simultaneously in one surface, giving rise to a trivial Berry phase when electrons encircling the two rings in the surface Brillouin zone. Since $\mathbb{G}_{x/y}$ only protect Dirac crossing of (010/100)-surface, (110)-, (1$\bar{1}$0)-, and ($\bar{1}10$)-surface will be gapped. Therefore, hinge states emerge at the intersection of (110)- and ($\bar{1}$10)-surface, via bending the (010)-surface along [010] direction~\cite{highorderfirst,Zhang136407,zhou2021glide}. Due to the existence of $\mathbb{G}_{x}$, gapped (110)- and ($\bar{1}$10)-surface will share opposite Dirac mass term, leading to nonsymmorphic symmetry protected hinge states as exhibited in Figure.~\ref{second}(j).

We then employ the Tl$_5$Te$_2$Br as an example to demonstrate the feasibility of BOFTI with third-order topology. For which, 10 Tl atoms occupy the 4c and 16l, 4 Te atoms occupy the 8h, and 2 Br atoms occupy the 4a~\cite{wyckoff, transport} as shown in Figures.~\ref{third}(a) and ~\ref{third}(b). The orbitally resolved band structures of Tl$_5$Te$_2$Br without and with SOC are displayed in Figures.~\ref{third}(c) and ~\ref{third}(d), respectively. In the absence of SOC, Tl-p and Te-p orbitals overlap and form a closed nodal line around the $\Gamma$ point, revealing a band inversion with the Tl-$p$ being lower than Te-$p$. The SOC gaps out the nodal line and drives the Tl$_5$Te$_2$Br to be an insulator with an energy gap of 56 meV. For this situation, a band inversion with opposite parities, i.e., different irreducible representations (irreps), is usually obtained and considered as a heuristic scenario of nontrivial bulk topology. However, the irreps of valence band maximum (VBM) at $\Gamma$ are the same with that of the conduction band minimum (CBM), resulting in a band inversion without irreps inversion and thus a trivial symmetry indicator. The valence bands can be decomposed into a linear combination of EBR, 
\begin{align}
	\begin{aligned}
		A_1@4a + &A_2@4a + E@4a + A_g@4c + 2A_1@8h + \\
		&B_1@8h + B_2@8h + A'@16l,
	\end{aligned}
\end{align}
suggesting the occupied outer-shell atomic orbitals of $s^2$, $s^2p^6$, and $s^2p^6$ for Tl, Te, and Br atoms, respectively. Therefore, the electrons of Tl-$p$ are transferred to both the Te-$p$ and Br-$p$ owing to the outer-shell atomic orbitals for respective elements are $s^2p^1$, $s^2p^5$ and $s^2p^4$. Interestingly, the electron transfer from 16l to 8h contributes an body-diagonal polarization, which may give rise to the BOFTI with third-order topology as discussed in our model analysis.

The fragile nature of Tl$_5$Te$_2$Br is identified by the Wilson loop spectra of isolated top four valence bands along a bent path as shown in Figure.~\ref{third}(e). Furthermore, Figures.~\ref{third}(f) -~\ref{third}(h) show us metallic surface states, exhibiting exotic wallpaper fermions such as hourglass fermion, fourfold-degenerate Dirac fermion, as a consequence of two perpendicular glide mirrors, $\mathbb{G}_{x/y} = \{\mathbb{M}_{x/y}|\frac{1}{2} \frac{1}{2} 0\}$. Similar to the case of TlGaTe$_2$, the Fermi surface of Tl$_5$Te$_2$Br still present us two rings with the left-handed and right-handed helicities in one surface, as exemplified in Figure.~S8~\cite{sm}. Then, we explore the corner states, which is the hallmark of a 3D third-order TI. To this end, we focus on a finite lattice composed of 6 $\times$ 6 $\times$ 6 conventional unit cell, and adopt the open boundary conditions for all the [100], [010], and [001] direction. As shown in Figure.~\ref{third}(i), 16 nearly degenerate in-gap states arise around the Fermi level. Notably, wave functions of the in-gap states are localized almost at 8 corners and vanishing in other region.
 
Inside the 454 materials of the TQC website who belong to space group $I4/mcm$ (No.140), eight of them are predicted to BOFTIs with surface states and higher-order topological phase, irrelative of SOC. Among them, five materials such as $X$GaTe$_2$ ($X$ = Tl, In, and Na), BaPbO$_3$ and InTe manifest metallic surface states on the (100)- and (010)- surface, exhibiting symmetry-protected hourglass fermion. In addition, the BOFTIs can coexist with the normal TI, which achieved in the InGaTe$_2$. As illustrated in Figure.~S10~\cite{sm}, while the surface states of normal TIs show up in both the top and bottom surfaces at the same time,  the surface states of BOFTIs can only emerge in one surface. Interestingly, along the intersection lines of (110)- and ($\bar{1}10$)-surface, two pairs of hinge states arise as a hallmark of SOTI. Moreover, three materials such as Tl$_5$Te$_2$Br and Tl$_4$Se$_3$ $Y$ ($Y$ = Pb and Sn) show surface states only on the (001)-surface with exotic hourglass fermion, and fourfold-degenerate Dirac fermion. The 16 nearly degenerate corner states only arise from the intersection of (100)-, (010)-, and (001)-surface. 
 
In conclusion, we demonstrate the coexistence of BOFTIs and higher-order topological phase with both the metallic surface states and hinge states or corner states. Remarkably, the emergence of surface states is a consequence of boundary obstruction, rather than the SOC, realizing the nontrivial characters in previously overlooked trivial insulators. With using TlGaTe$_2$ and Tl$_5$Te$_2$Br as two examples, we show that tangent and body-diagonal boundary polarizations result in the birth of second- and third-order TIs, respectively, as identified by the energy spectrum calculations of nanorod and finite lattice. Especially, TlGaTe$_2$ exhibits a bulk energy gap more than 1 eV, which provides advantages for experiments to detect obstructed boundary states on the surface and edge. 
Our study lay the groundwork of finding exotic topological phases through boundary obstruction, and promote it as a new platform for exploring the intriguing physics of exotic surface states and fragile topological phases.

\section*{ACKNOWLEDGEMENTS}
We thank helpful discussions with Yuanfeng Xu and B. A. Bernevig at Princeton University. This work was supported by the National Natural Science Foundation of China (Grants No. 11904205, No. 12074217, and No. 12174220), the Shandong Provincial Natural Science Foundation of China (Grants No. ZR2019QA019 and No. ZR2019MEM013), the Shandong Provincial Key Research and Development Program (Major Scientific and Technological Innovation Project) (Grant No. 2019JZZY010302), and the Qilu Young Scholar Program of Shandong University. B.Y. acknowledges the financial support by the European Research Council (ERC Consolidator Grant ``NonlinearTopo'', No. 815869) and the Max Planck Lab.

\section*{COMPETING INTERESTS}
The authors declare no competing interests.

\section*{DATA AVAILABILITY}
The data that support the ﬁndings of this study are available from the corresponding author upon reasonable request.


\begin{thebibliography}{10}
	\expandafter\ifx\csname url\endcsname\relax
	\def\url#1{\texttt{#1}}\fi
	\expandafter\ifx\csname urlprefix\endcsname\relax\def\urlprefix{URL }\fi
	\providecommand{\bibinfo}[2]{#2}
	\providecommand{\eprint}[2][]{\url{#2}}
	
	\bibitem{boundary}
	\bibinfo{author}{X.~Wu}, \bibinfo{author}{W.~A. Benalcazar},
	\bibinfo{author}{Y.~Li}, \bibinfo{author}{R.~Thomale}, \bibinfo{author}{C.-X.
		Liu}, \bibinfo{author}{J.~Hu}.
	\newblock \bibinfo{title}{Boundary-obstructed topological
		high-${\mathit{t}}_{c}$ superconductivity in iron pnictides}.
	\newblock \emph{\bibinfo{journal}{Phys. Rev. X}} \textbf{\bibinfo{volume}{10}},
	\bibinfo{pages}{041014} (\bibinfo{year}{2020}).
	
	\bibitem{superboundary}
	\bibinfo{author}{A.~Tiwari}, \bibinfo{author}{A.~Jahin},
	\bibinfo{author}{Y.~Wang}.
	\newblock \bibinfo{title}{Chiral dirac superconductors: Second-order and
		boundary-obstructed topology}.
	\newblock \emph{\bibinfo{journal}{Phys. Rev. Res.}}
	\textbf{\bibinfo{volume}{2}}, \bibinfo{pages}{043300} (\bibinfo{year}{2020}).
	
	\bibitem{ezawaboundary}
	\bibinfo{author}{M.~Ezawa}.
	\newblock \bibinfo{title}{Edge-corner correspondence: Boundary-obstructed
		topological phases with chiral symmetry}.
	\newblock \emph{\bibinfo{journal}{Phys. Rev. B}}
	\textbf{\bibinfo{volume}{102}}, \bibinfo{pages}{121405}
	(\bibinfo{year}{2020}).
	
	\bibitem{khalaf2021boundary}
	\bibinfo{author}{E.~Khalaf}, \bibinfo{author}{W.~A. Benalcazar},
	\bibinfo{author}{T.~L. Hughes}, \bibinfo{author}{R.~Queiroz}.
	\newblock \bibinfo{title}{Boundary-obstructed topological phases}.
	\newblock \emph{\bibinfo{journal}{Phys. Rev. Res.}}
	\textbf{\bibinfo{volume}{3}}, \bibinfo{pages}{013239} (\bibinfo{year}{2021}).
	
	\bibitem{hasan22}
	\bibinfo{author}{M.~Z. Hasan}, \bibinfo{author}{C.~L. Kane}.
	\newblock \bibinfo{title}{Colloquium: topological insulators}.
	\newblock \emph{\bibinfo{journal}{Rev. Mod. Phys.}}
	\textbf{\bibinfo{volume}{82}}, \bibinfo{pages}{3045} (\bibinfo{year}{2010}).
	
	\bibitem{qi2011topological}
	\bibinfo{author}{X.-L. Qi}, \bibinfo{author}{S.-C. Zhang}.
	\newblock \bibinfo{title}{Topological insulators and superconductors}.
	\newblock \emph{\bibinfo{journal}{Rev. Mod. Phys.}}
	\textbf{\bibinfo{volume}{83}}, \bibinfo{pages}{1057} (\bibinfo{year}{2011}).
	
	\bibitem{Bansil2016rmp}
	\bibinfo{author}{A.~Bansil}, \bibinfo{author}{H.~Lin},
	\bibinfo{author}{T.~Das}.
	\newblock \bibinfo{title}{Colloquium: Topological band theory}.
	\newblock \emph{\bibinfo{journal}{Rev. Mod. Phys.}}
	\textbf{\bibinfo{volume}{88}}, \bibinfo{pages}{021004}
	(\bibinfo{year}{2016}).
	
	\bibitem{Ando2015}
	\bibinfo{author}{Y.~Ando}, \bibinfo{author}{L.~Fu}.
	\newblock \bibinfo{title}{{Topological Crystalline Insulators and Topological
			Superconductors: From Concepts to Materials}}.
	\newblock \emph{\bibinfo{journal}{Annu. Rev. Condens. Matter Phys.}}
	\textbf{\bibinfo{volume}{6}}, \bibinfo{pages}{361} (\bibinfo{year}{2015}).
	
	\bibitem{Xiao2021NRP}
	\bibinfo{author}{J.~Xiao}, \bibinfo{author}{B.~Yan}.
	\newblock \bibinfo{title}{{First- principles calculations for topological
			quantum materials}}.
	\newblock \emph{\bibinfo{journal}{Nat. Rev. Phys.}}
	\textbf{\bibinfo{volume}{3}}, \bibinfo{pages}{283--297}
	(\bibinfo{year}{2021}).
	
	\bibitem{tqc1}
	\bibinfo{author}{B.~Bradlyn}, \bibinfo{author}{L.~Elcoro},
	\bibinfo{author}{J.~Cano}, \bibinfo{author}{M.~G. Vergniory},
	\bibinfo{author}{Z.~Wang}, \bibinfo{author}{C.~Felser},
	\bibinfo{author}{M.~I. Aroyo}, \bibinfo{author}{B.~A. Bernevig}.
	\newblock \bibinfo{title}{{Topological quantum chemistry}}.
	\newblock \emph{\bibinfo{journal}{Nature}} \textbf{\bibinfo{volume}{547}},
	\bibinfo{pages}{298--305} (\bibinfo{year}{2017}).
	
	\bibitem{Po2017}
	\bibinfo{author}{H.~C. Po}, \bibinfo{author}{A.~Vishwanath},
	\bibinfo{author}{H.~Watanabe}.
	\newblock \bibinfo{title}{{Symmetry-based Indicators of Band Topology in the
			230 Space Groups}}.
	\newblock \emph{\bibinfo{journal}{Nat. Commun.}} \textbf{\bibinfo{volume}{8}},
	\bibinfo{pages}{50} (\bibinfo{year}{2017}).
	
	\bibitem{Song2018}
	\bibinfo{author}{Z.~Song}, \bibinfo{author}{T.~Zhang},
	\bibinfo{author}{Z.~Fang}, \bibinfo{author}{C.~Fang}.
	\newblock \bibinfo{title}{{Quantitative mappings between symmetry and topology
			in solids}}.
	\newblock \emph{\bibinfo{journal}{Nat. Commun.}} \textbf{\bibinfo{volume}{9}},
	\bibinfo{pages}{3530} (\bibinfo{year}{2018}).
	
	\bibitem{Watanabe2018}
	\bibinfo{author}{H.~Watanabe}, \bibinfo{author}{H.~C. Po},
	\bibinfo{author}{A.~Vishwanath}.
	\newblock \bibinfo{title}{{Structure and topology of band structures in the
			1651 magnetic space groups}}.
	\newblock \emph{\bibinfo{journal}{Sci. Adv.}} \textbf{\bibinfo{volume}{4}},
	\bibinfo{pages}{eaat8685} (\bibinfo{year}{2018}).
	
	\bibitem{Tang2019}
	\bibinfo{author}{F.~Tang}, \bibinfo{author}{H.~C. Po},
	\bibinfo{author}{A.~Vishwanath}, \bibinfo{author}{X.~Wan}.
	\newblock \bibinfo{title}{Topological materials discovery by large-order
		symmetry indicators}.
	\newblock \emph{\bibinfo{journal}{Sci. Adv.}} \textbf{\bibinfo{volume}{5}},
	\bibinfo{pages}{eaau8725} (\bibinfo{year}{2019}).
	
	\bibitem{nature1}
	\bibinfo{author}{M.~Vergniory}, \bibinfo{author}{L.~Elcoro},
	\bibinfo{author}{C.~Felser}, \bibinfo{author}{N.~Regnault},
	\bibinfo{author}{B.~A. Bernevig}, \bibinfo{author}{Z.~Wang}.
	\newblock \bibinfo{title}{A complete catalogue of high-quality topological
		materials}.
	\newblock \emph{\bibinfo{journal}{Nature}} \textbf{\bibinfo{volume}{566}},
	\bibinfo{pages}{480--485} (\bibinfo{year}{2019}).
	
	\bibitem{nature2}
	\bibinfo{author}{T.~Zhang}, \bibinfo{author}{Y.~Jiang},
	\bibinfo{author}{Z.~Song}, \bibinfo{author}{H.~Huang},
	\bibinfo{author}{Y.~He}, \bibinfo{author}{Z.~Fang},
	\bibinfo{author}{H.~Weng}, \bibinfo{author}{C.~Fang}.
	\newblock \bibinfo{title}{Catalogue of topological electronic materials}.
	\newblock \emph{\bibinfo{journal}{Nature}} \textbf{\bibinfo{volume}{566}},
	\bibinfo{pages}{475--479} (\bibinfo{year}{2019}).
	
	\bibitem{nature3}
	\bibinfo{author}{F.~Tang}, \bibinfo{author}{H.~C. Po},
	\bibinfo{author}{A.~Vishwanath}, \bibinfo{author}{X.~Wan}.
	\newblock \bibinfo{title}{Comprehensive search for topological materials using
		symmetry indicators}.
	\newblock \emph{\bibinfo{journal}{Nature}} \textbf{\bibinfo{volume}{566}},
	\bibinfo{pages}{486--489} (\bibinfo{year}{2019}).
	
	\bibitem{xu2021three}
	\bibinfo{author}{Y.~Xu}, \bibinfo{author}{L.~Elcoro}, \bibinfo{author}{G.~Li},
	\bibinfo{author}{Z.-D. Song}, \bibinfo{author}{N.~Regnault},
	\bibinfo{author}{Q.~Yang}, \bibinfo{author}{Y.~Sun},
	\bibinfo{author}{S.~Parkin}, \bibinfo{author}{C.~Felser},
	\bibinfo{author}{B.~A. Bernevig}.
	\newblock \bibinfo{title}{Three-dimensional real space invariants, obstructed
		atomic insulators and a new principle for active catalytic sites}.
	\newblock \emph{\bibinfo{journal}{arXiv:2111.02433}}  (\bibinfo{year}{2021}).
	
	\bibitem{xu2021filling}
	\bibinfo{author}{Y.~Xu}, \bibinfo{author}{L.~Elcoro}, \bibinfo{author}{Z.-D.
		Song}, \bibinfo{author}{M.~Vergniory}, \bibinfo{author}{C.~Felser},
	\bibinfo{author}{S.~S. Parkin}, \bibinfo{author}{N.~Regnault},
	\bibinfo{author}{J.~L. Ma{\~n}es}, \bibinfo{author}{B.~A. Bernevig}.
	\newblock \bibinfo{title}{Filling-enforced obstructed atomic insulators}.
	\newblock \emph{\bibinfo{journal}{arXiv:2106.10276}}  (\bibinfo{year}{2021}).
	
	\bibitem{Nelson2021PRL}
	\bibinfo{author}{A.~Nelson}, \bibinfo{author}{T.~Neupert},
	\bibinfo{author}{T.~c.~v. Bzdu\ifmmode~\check{s}\else \v{s}\fi{}ek},
	\bibinfo{author}{A.~Alexandradinata}.
	\newblock \bibinfo{title}{Multicellularity of delicate topological insulators}.
	\newblock \emph{\bibinfo{journal}{Phys. Rev. Lett.}}
	\textbf{\bibinfo{volume}{126}}, \bibinfo{pages}{216404}
	(\bibinfo{year}{2021}).
	
	\bibitem{schindler2021non}
	\bibinfo{author}{F.~Schindler}, \bibinfo{author}{B.~A. Bernevig}.
	\newblock \bibinfo{title}{Noncompact atomic insulators}.
	\newblock \emph{\bibinfo{journal}{Phys. Rev. B}}
	\textbf{\bibinfo{volume}{104}}, \bibinfo{pages}{L201114}
	(\bibinfo{year}{2021}).
	
	\bibitem{unconventional}
	\bibinfo{author}{J.~Gao}, \bibinfo{author}{Y.~Qian}, \bibinfo{author}{H.~Jia},
	\bibinfo{author}{Z.~Guo}, \bibinfo{author}{Z.~Fang},
	\bibinfo{author}{M.~Liu}, \bibinfo{author}{H.~Weng},
	\bibinfo{author}{Z.~Wang}.
	\newblock \bibinfo{title}{Unconventional materials: the mismatch between
		electronic charge centers and atomic positions}.
	\newblock \emph{\bibinfo{journal}{Sci. Bull.}}  (\bibinfo{year}{2022}).
	
	\bibitem{oaicatalytic}
	\bibinfo{author}{G.~Li}, \bibinfo{author}{Y.~Xu}, \bibinfo{author}{Z.~Song},
	\bibinfo{author}{Q.~Yang}, \bibinfo{author}{U.~Gupta},
	\bibinfo{author}{Y.~Sun}, \bibinfo{author}{P.~Sessi}, \bibinfo{author}{S.~S.
		Parkin}, \bibinfo{author}{B.~A. Bernevig}, \bibinfo{author}{C.~Felser},
	et~al.
	\newblock \bibinfo{title}{Obstructed surface states as the origin of catalytic
		activity in inorganic heterogeneous catalysts}.
	\newblock \emph{\bibinfo{journal}{arXiv:2111.02435}}  (\bibinfo{year}{2021}).
	
	\bibitem{Benalcazar61}
	\bibinfo{author}{W.~A. Benalcazar}, \bibinfo{author}{B.~A. Bernevig},
	\bibinfo{author}{T.~L. Hughes}.
	\newblock \bibinfo{title}{Quantized electric multipole insulators}.
	\newblock \emph{\bibinfo{journal}{Science}} \textbf{\bibinfo{volume}{357}},
	\bibinfo{pages}{61--66} (\bibinfo{year}{2017}).
	
	\bibitem{highorderfirst}
	\bibinfo{author}{F.~Schindler}, \bibinfo{author}{A.~M. Cook},
	\bibinfo{author}{M.~G. Vergniory}, \bibinfo{author}{Z.~Wang},
	\bibinfo{author}{S.~S. Parkin}, \bibinfo{author}{B.~A. Bernevig},
	\bibinfo{author}{T.~Neupert}.
	\newblock \bibinfo{title}{Higher-order topological insulators}.
	\newblock \emph{\bibinfo{journal}{Sci. Adv.}} \textbf{\bibinfo{volume}{4}},
	\bibinfo{pages}{eaat0346} (\bibinfo{year}{2018}).
	
	\bibitem{frankbi}
	\bibinfo{author}{F.~Schindler}, \bibinfo{author}{Z.~Wang},
	\bibinfo{author}{M.~G. Vergniory}, \bibinfo{author}{A.~M. Cook},
	\bibinfo{author}{A.~Murani}, \bibinfo{author}{S.~Sengupta},
	\bibinfo{author}{A.~Y. Kasumov}, \bibinfo{author}{R.~Deblock},
	\bibinfo{author}{S.~Jeon}, \bibinfo{author}{I.~Drozdov},
	\bibinfo{author}{H.~Bouchiat}, \bibinfo{author}{S.~Gu{\'{e}}ron},
	\bibinfo{author}{A.~Yazdani}, \bibinfo{author}{B.~A. Bernevig},
	\bibinfo{author}{T.~Neupert}.
	\newblock \bibinfo{title}{{Higher-order topology in bismuth}}.
	\newblock \emph{\bibinfo{journal}{Nat. Phys.}} \textbf{\bibinfo{volume}{14}},
	\bibinfo{pages}{918--924} (\bibinfo{year}{2018}).
	
	\bibitem{higherordereuln2as2}
	\bibinfo{author}{Y.~Xu}, \bibinfo{author}{Z.~Song}, \bibinfo{author}{Z.~Wang},
	\bibinfo{author}{H.~Weng}, \bibinfo{author}{X.~Dai}.
	\newblock \bibinfo{title}{Higher-order topology of the axion insulator
		{EuIn$_2$As$_2$}}.
	\newblock \emph{\bibinfo{journal}{Phys. Rev. Lett.}}
	\textbf{\bibinfo{volume}{122}}, \bibinfo{pages}{256402}
	(\bibinfo{year}{2019}).
	
	\bibitem{highorderbieuo}
	\bibinfo{author}{C.~Chen}, \bibinfo{author}{Z.~Song}, \bibinfo{author}{J.-Z.
		Zhao}, \bibinfo{author}{Z.~Chen}, \bibinfo{author}{Z.-M. Yu},
	\bibinfo{author}{X.-L. Sheng}, \bibinfo{author}{S.~A. Yang}.
	\newblock \bibinfo{title}{Universal approach to magnetic second-order
		topological insulator}.
	\newblock \emph{\bibinfo{journal}{Phys. Rev. Lett.}}
	\textbf{\bibinfo{volume}{125}}, \bibinfo{pages}{056402}
	(\bibinfo{year}{2020}).
	
	\bibitem{Yue2019}
	\bibinfo{author}{C.~Yue}, \bibinfo{author}{Y.~Xu}, \bibinfo{author}{Z.~Song},
	\bibinfo{author}{H.~Weng}, \bibinfo{author}{Y.~M. Lu},
	\bibinfo{author}{C.~Fang}, \bibinfo{author}{X.~Dai}.
	\newblock \bibinfo{title}{{Symmetry-enforced chiral hinge states and surface
			quantum anomalous Hall effect in the magnetic axion insulator
			Bi$_{2-x}$Sm$_x$Se$_3$}}.
	\newblock \emph{\bibinfo{journal}{Nat. Phys.}} \textbf{\bibinfo{volume}{15}},
	\bibinfo{pages}{577--581} (\bibinfo{year}{2019}).
	
	\bibitem{Park216803}
	\bibinfo{author}{M.~J. Park}, \bibinfo{author}{Y.~Kim}, \bibinfo{author}{G.~Y.
		Cho}, \bibinfo{author}{S.~Lee}.
	\newblock \bibinfo{title}{Higher-order topological insulator in twisted bilayer
		graphene}.
	\newblock \emph{\bibinfo{journal}{Phys. Rev. Lett.}}
	\textbf{\bibinfo{volume}{123}}, \bibinfo{pages}{216803}
	(\bibinfo{year}{2019}).
	
	\bibitem{kempkes2019robust}
	\bibinfo{author}{S.~Kempkes}, \bibinfo{author}{M.~Slot},
	\bibinfo{author}{J.~van Den~Broeke}, \bibinfo{author}{P.~Capiod},
	\bibinfo{author}{W.~Benalcazar}, \bibinfo{author}{D.~Vanmaekelbergh},
	\bibinfo{author}{D.~Bercioux}, \bibinfo{author}{I.~Swart},
	\bibinfo{author}{C.~M. Smith}.
	\newblock \bibinfo{title}{Robust zero-energy modes in an electronic
		higher-order topological insulator}.
	\newblock \emph{\bibinfo{journal}{Nat. Mater.}} \textbf{\bibinfo{volume}{18}},
	\bibinfo{pages}{1292--1297} (\bibinfo{year}{2019}).
	
	\bibitem{highordergranpj}
	\bibinfo{author}{E.~Lee}, \bibinfo{author}{R.~Kim}, \bibinfo{author}{J.~Ahn},
	\bibinfo{author}{B.-J. Yang}.
	\newblock \bibinfo{title}{Two-dimensional higher-order topology in monolayer
		graphdiyne}.
	\newblock \emph{\bibinfo{journal}{npj Quantum Mater.}}
	\textbf{\bibinfo{volume}{5}}, \bibinfo{pages}{1--7} (\bibinfo{year}{2020}).
	
	\bibitem{highordergraprl}
	\bibinfo{author}{X.-L. Sheng}, \bibinfo{author}{C.~Chen},
	\bibinfo{author}{H.~Liu}, \bibinfo{author}{Z.~Chen}, \bibinfo{author}{Z.-M.
		Yu}, \bibinfo{author}{Y.~Zhao}, \bibinfo{author}{S.~A. Yang}.
	\newblock \bibinfo{title}{Two-dimensional second-order topological insulator in
		graphdiyne}.
	\newblock \emph{\bibinfo{journal}{Phys. Rev. Lett.}}
	\textbf{\bibinfo{volume}{123}}, \bibinfo{pages}{256402}
	(\bibinfo{year}{2019}).
	
	\bibitem{highordersplit}
	\bibinfo{author}{B.~Liu}, \bibinfo{author}{G.~Zhao}, \bibinfo{author}{Z.~Liu},
	\bibinfo{author}{Z.~Wang}.
	\newblock \bibinfo{title}{Two-dimensional quadrupole topological insulator in
		$\gamma$-graphyne}.
	\newblock \emph{\bibinfo{journal}{Nano Lett.}} \textbf{\bibinfo{volume}{19}},
	\bibinfo{pages}{6492--6497} (\bibinfo{year}{2019}).
	
	\bibitem{Garcia2018}
	\bibinfo{author}{M.~Serra-Garcia}, \bibinfo{author}{V.~Peri},
	\bibinfo{author}{R.~S{\"{u}}sstrunk}, \bibinfo{author}{O.~R. Bilal},
	\bibinfo{author}{T.~Larsen}, \bibinfo{author}{L.~G. Villanueva},
	\bibinfo{author}{S.~D. Huber}.
	\newblock \bibinfo{title}{{Observation of a phononic quadrupole topological
			insulator}}.
	\newblock \emph{\bibinfo{journal}{Nature}} \textbf{\bibinfo{volume}{555}},
	\bibinfo{pages}{342--345} (\bibinfo{year}{2018}).
	
	\bibitem{Imhof2018}
	\bibinfo{author}{S.~Imhof}, \bibinfo{author}{C.~Berger},
	\bibinfo{author}{F.~Bayer}, \bibinfo{author}{J.~Brehm},
	\bibinfo{author}{L.~W. Molenkamp}, \bibinfo{author}{T.~Kiessling},
	\bibinfo{author}{F.~Schindler}, \bibinfo{author}{C.~H. Lee},
	\bibinfo{author}{M.~Greiter}, \bibinfo{author}{T.~Neupert},
	\bibinfo{author}{R.~Thomale}.
	\newblock \bibinfo{title}{{Topolectrical-circuit realization of topological
			corner modes}}.
	\newblock \emph{\bibinfo{journal}{Nat. Phys.}} \textbf{\bibinfo{volume}{14}},
	\bibinfo{pages}{925--929} (\bibinfo{year}{2018}).
	
	\bibitem{Peterson2018}
	\bibinfo{author}{C.~W. Peterson}, \bibinfo{author}{W.~A. Benalcazar},
	\bibinfo{author}{T.~L. Hughes}, \bibinfo{author}{G.~Bahl}.
	\newblock \bibinfo{title}{{A quantized microwave quadrupole insulator with
			topologically protected corner states}}.
	\newblock \emph{\bibinfo{journal}{Nature}} \textbf{\bibinfo{volume}{555}},
	\bibinfo{pages}{346--350} (\bibinfo{year}{2018}).
	
	\bibitem{xue2019acoustic}
	\bibinfo{author}{H.~Xue}, \bibinfo{author}{Y.~Yang}, \bibinfo{author}{F.~Gao},
	\bibinfo{author}{Y.~Chong}, \bibinfo{author}{B.~Zhang}.
	\newblock \bibinfo{title}{Acoustic higher-order topological insulator on a
		kagome lattice}.
	\newblock \emph{\bibinfo{journal}{Nat. Mater.}} \textbf{\bibinfo{volume}{18}},
	\bibinfo{pages}{108--112} (\bibinfo{year}{2019}).
	
	\bibitem{ni2019observation}
	\bibinfo{author}{X.~Ni}, \bibinfo{author}{M.~Weiner}, \bibinfo{author}{A.~Alu},
	\bibinfo{author}{A.~B. Khanikaev}.
	\newblock \bibinfo{title}{Observation of higher-order topological acoustic
		states protected by generalized chiral symmetry}.
	\newblock \emph{\bibinfo{journal}{Nat. Mater.}} \textbf{\bibinfo{volume}{18}},
	\bibinfo{pages}{113--120} (\bibinfo{year}{2019}).
	
	\bibitem{noguchi2021evidence}
	\bibinfo{author}{R.~Noguchi}, \bibinfo{author}{M.~Kobayashi},
	\bibinfo{author}{Z.~Jiang}, \bibinfo{author}{K.~Kuroda},
	\bibinfo{author}{T.~Takahashi}, \bibinfo{author}{Z.~Xu},
	\bibinfo{author}{D.~Lee}, \bibinfo{author}{M.~Hirayama},
	\bibinfo{author}{M.~Ochi}, \bibinfo{author}{T.~Shirasawa}, et~al.
	\newblock \bibinfo{title}{Evidence for a higher-order topological insulator in
		a three-dimensional material built from van der waals stacking of
		bismuth-halide chains}.
	\newblock \emph{\bibinfo{journal}{Nat. Mater.}} \textbf{\bibinfo{volume}{20}},
	\bibinfo{pages}{473--479} (\bibinfo{year}{2021}).
	
	\bibitem{multipolescience}
	\bibinfo{author}{W.~A. Benalcazar}, \bibinfo{author}{B.~A. Bernevig},
	\bibinfo{author}{T.~L. Hughes}.
	\newblock \bibinfo{title}{Quantized electric multipole insulators}.
	\newblock \emph{\bibinfo{journal}{Science}} \textbf{\bibinfo{volume}{357}},
	\bibinfo{pages}{61--66} (\bibinfo{year}{2017}).
	
	\bibitem{benalcazar2017electric}
	\bibinfo{author}{W.~A. Benalcazar}, \bibinfo{author}{B.~A. Bernevig},
	\bibinfo{author}{T.~L. Hughes}.
	\newblock \bibinfo{title}{Electric multipole moments, topological multipole
		moment pumping, and chiral hinge states in crystalline insulators}.
	\newblock \emph{\bibinfo{journal}{Phys. Rev. B}} \textbf{\bibinfo{volume}{96}},
	\bibinfo{pages}{245115} (\bibinfo{year}{2017}).
	
	\bibitem{tqc3}
	\bibinfo{author}{J.~Cano}, \bibinfo{author}{B.~Bradlyn},
	\bibinfo{author}{Z.~Wang}, \bibinfo{author}{L.~Elcoro},
	\bibinfo{author}{M.~Vergniory}, \bibinfo{author}{C.~Felser},
	\bibinfo{author}{M.~Aroyo}, \bibinfo{author}{B.~A. Bernevig}.
	\newblock \bibinfo{title}{Topology of disconnected elementary band
		representations}.
	\newblock \emph{\bibinfo{journal}{Phys. Rev. Lett.}}
	\textbf{\bibinfo{volume}{120}}, \bibinfo{pages}{266401}
	(\bibinfo{year}{2018}).
	
	\bibitem{cano2021topology}
	\bibinfo{author}{J.~Cano}, \bibinfo{author}{L.~Elcoro},
	\bibinfo{author}{M.~Aroyo}, \bibinfo{author}{B.~A. Bernevig},
	\bibinfo{author}{B.~Bradlyn}.
	\newblock \bibinfo{title}{Topology invisible to eigenvalues in obstructed
		atomic insulators}.
	\newblock \emph{\bibinfo{journal}{arXiv:2107.00647}}  (\bibinfo{year}{2021}).
	
	\bibitem{song2020twisted}
	\bibinfo{author}{Z.-D. Song}, \bibinfo{author}{L.~Elcoro},
	\bibinfo{author}{B.~A. Bernevig}.
	\newblock \bibinfo{title}{Twisted bulk-boundary correspondence of fragile
		topology}.
	\newblock \emph{\bibinfo{journal}{Science}} \textbf{\bibinfo{volume}{367}},
	\bibinfo{pages}{794--797} (\bibinfo{year}{2020}).
	
	\bibitem{po2017symmetry}
	\bibinfo{author}{H.~C. Po}, \bibinfo{author}{A.~Vishwanath},
	\bibinfo{author}{H.~Watanabe}.
	\newblock \bibinfo{title}{Symmetry-based indicators of band topology in the 230
		space groups}.
	\newblock \emph{\bibinfo{journal}{Nat. Commun.}} \textbf{\bibinfo{volume}{8}},
	\bibinfo{pages}{1--9} (\bibinfo{year}{2017}).
	
	\bibitem{sm}
	\emph{\bibinfo{journal}{See Supplemental Material (i) calculation methods, (ii)
			model results about the relation between boundary polarizations and
			higher-order topological phases, (iii) character table of point group $m
			(Cs)$ and $mm2 (C2v)$, (iv) band structures, Wilson loop spectra and energy
			level of finite lattice for the corresponding normal insulator, TI, SOTI, and
			TOTI, (v) the edge states and Wilson loop spectra of TlGaTe$_2$ and
			Tl$_5$Te$_2$Br, (vi) the geometry structures, bulk bands, Wilson bands,
			surface bands and nanorod bands of the SOTI material candidates, (vii) the
			geometry structures, bulk bands, Wilson bands, surface bands, energy level of
			finite lattice and probability of corner states of TOTI material candidates,
			which includes Refs. [58–72]}} .
	
	\bibitem{wyckoff}
	\bibinfo{author}{M.~I. Aroyo}, \bibinfo{author}{J.~M. Perez-Mato},
	\bibinfo{author}{C.~Capillas}, \bibinfo{author}{E.~Kroumova},
	\bibinfo{author}{S.~Ivantchev}, \bibinfo{author}{G.~Madariaga},
	\bibinfo{author}{A.~Kirov}, \bibinfo{author}{H.~Wondratschek}.
	\newblock \bibinfo{title}{Bilbao crystallographic server: I. databases and
		crystallographic computing programs}.
	\newblock \emph{\bibinfo{journal}{Z. Kristallogr. Cryst. Mater.}}
	\textbf{\bibinfo{volume}{221}}, \bibinfo{pages}{15--27}
	(\bibinfo{year}{2006}).
	
	\bibitem{transport}
	\bibinfo{author}{F.~Ricci}, \bibinfo{author}{W.~Chen},
	\bibinfo{author}{U.~Aydemir}, \bibinfo{author}{G.~J. Snyder},
	\bibinfo{author}{G.-M. Rignanese}, \bibinfo{author}{A.~Jain},
	\bibinfo{author}{G.~Hautier}.
	\newblock \bibinfo{title}{An ab initio electronic transport database for
		inorganic materials}.
	\newblock \emph{\bibinfo{journal}{Sci. Data}} \textbf{\bibinfo{volume}{4}},
	\bibinfo{pages}{1--13} (\bibinfo{year}{2017}).
	
	\bibitem{hse06}
	\bibinfo{author}{A.~V. Krukau}, \bibinfo{author}{O.~A. Vydrov},
	\bibinfo{author}{A.~F. Izmaylov}, \bibinfo{author}{G.~E. Scuseria}.
	\newblock \bibinfo{title}{Influence of the exchange screening parameter on the
		performance of screened hybrid functionals}.
	\newblock \emph{\bibinfo{journal}{J. Chem. Phys.}}
	\textbf{\bibinfo{volume}{125}}, \bibinfo{pages}{224106}
	(\bibinfo{year}{2006}).
	
	\bibitem{wcc}
	\bibinfo{author}{R.~Yu}, \bibinfo{author}{X.~L. Qi},
	\bibinfo{author}{A.~Bernevig}, \bibinfo{author}{Z.~Fang},
	\bibinfo{author}{X.~Dai}.
	\newblock \bibinfo{title}{Equivalent expression of {Z$_2$} topological
		invariant for band insulators using the non-abelian berry connection}.
	\newblock \emph{\bibinfo{journal}{Phys. Rev. B}} \textbf{\bibinfo{volume}{84}},
	\bibinfo{pages}{075119} (\bibinfo{year}{2011}).
	
	\bibitem{fragile3}
	\bibinfo{author}{Z.-D. Song}, \bibinfo{author}{L.~Elcoro},
	\bibinfo{author}{Y.-F. Xu}, \bibinfo{author}{N.~Regnault},
	\bibinfo{author}{B.~A. Bernevig}.
	\newblock \bibinfo{title}{Fragile phases as affine monoids: Classification and
		material examples}.
	\newblock \emph{\bibinfo{journal}{Phys. Rev. X}} \textbf{\bibinfo{volume}{10}},
	\bibinfo{pages}{031001} (\bibinfo{year}{2020}).
	
	\bibitem{fragile4}
	\bibinfo{author}{A.~Luo}, \bibinfo{author}{Z.~Song}, \bibinfo{author}{G.~Xu}.
	\newblock \bibinfo{title}{Fragile topological band in the checkerboard
		antiferromagnetic monolayer fese}.
	\newblock \emph{\bibinfo{journal}{npj Comput. Mater.}}
	\textbf{\bibinfo{volume}{8}}, \bibinfo{pages}{26} (\bibinfo{year}{2022}).
	
	\bibitem{fragile1}
	\bibinfo{author}{H.~C. Po}, \bibinfo{author}{H.~Watanabe},
	\bibinfo{author}{A.~Vishwanath}.
	\newblock \bibinfo{title}{Fragile topology and wannier obstructions}.
	\newblock \emph{\bibinfo{journal}{Phys. Rev. Lett.}}
	\textbf{\bibinfo{volume}{121}}, \bibinfo{pages}{126402}
	(\bibinfo{year}{2018}).
	
	\bibitem{fragile2}
	\bibinfo{author}{B.~Bradlyn}, \bibinfo{author}{Z.~Wang},
	\bibinfo{author}{J.~Cano}, \bibinfo{author}{B.~A. Bernevig}.
	\newblock \bibinfo{title}{Disconnected elementary band representations, fragile
		topology, and wilson loops as topological indices: An example on the
		triangular lattice}.
	\newblock \emph{\bibinfo{journal}{Phys. Rev. B}} \textbf{\bibinfo{volume}{99}},
	\bibinfo{pages}{045140} (\bibinfo{year}{2019}).
	
	\bibitem{Zhang136407}
	\bibinfo{author}{R.-X. Zhang}, \bibinfo{author}{F.~Wu},
	\bibinfo{author}{S.~Das~Sarma}.
	\newblock \bibinfo{title}{M\"obius insulator and higher-order topology in
		{MnBi$_{2n}$Te$_{3n+1}$}}.
	\newblock \emph{\bibinfo{journal}{Phys. Rev. Lett.}}
	\textbf{\bibinfo{volume}{124}}, \bibinfo{pages}{136407}
	(\bibinfo{year}{2020}).
	
	\bibitem{zhou2021glide}
	\bibinfo{author}{X.~Zhou}, \bibinfo{author}{C.-H. Hsu}, \bibinfo{author}{C.-Y.
		Huang}, \bibinfo{author}{M.~Iraola}, \bibinfo{author}{J.~L. Ma{\~n}es},
	\bibinfo{author}{M.~G. Vergniory}, \bibinfo{author}{H.~Lin},
	\bibinfo{author}{N.~Kioussis}.
	\newblock \bibinfo{title}{Glide symmetry protected higher-order topological
		insulators from semimetals with butterfly-like nodal lines}.
	\newblock \emph{\bibinfo{journal}{npj Comput. Mater.}}
	\textbf{\bibinfo{volume}{7}}, \bibinfo{pages}{1--6} (\bibinfo{year}{2021}).
	
	\bibitem{pbevasp}
	\bibinfo{author}{J.~P. Perdew}, \bibinfo{author}{K.~Burke},
	\bibinfo{author}{M.~Ernzerhof}.
	\newblock \bibinfo{title}{Generalized gradient approximation made simple}.
	\newblock \emph{\bibinfo{journal}{Phys. Rev. Lett.}}
	\textbf{\bibinfo{volume}{77}}, \bibinfo{pages}{3865} (\bibinfo{year}{1996}).
	
	\bibitem{vasp}
	\bibinfo{author}{G.~Kresse}, \bibinfo{author}{J.~Furthm{\"{u}}ller}.
	\newblock \bibinfo{title}{{Efficient iterative schemes for ab initio
			total-energy calculations using a plane-wave basis set}}.
	\newblock \emph{\bibinfo{journal}{Phys. Rev. B}} \textbf{\bibinfo{volume}{54}},
	\bibinfo{pages}{11169--11186} (\bibinfo{year}{1996}).
	
	\bibitem{fleur}
	
	\newblock \bibinfo{howpublished}{See \url{http://www.flapw.de}}.
	
	\bibitem{wannier90}
	\bibinfo{author}{G.~Pizzi}, \bibinfo{author}{V.~Vitale},
	\bibinfo{author}{R.~Arita}, \bibinfo{author}{S.~Bl{\"u}gel},
	\bibinfo{author}{F.~Freimuth}, \bibinfo{author}{G.~G{\'e}ranton},
	\bibinfo{author}{M.~Gibertini}, \bibinfo{author}{D.~Gresch},
	\bibinfo{author}{C.~Johnson}, \bibinfo{author}{T.~Koretsune}, et~al.
	\newblock \bibinfo{title}{Wannier90 as a community code: {New} features and
		applications}.
	\newblock \emph{\bibinfo{journal}{J. Phys. Condens. Matter}}
	\textbf{\bibinfo{volume}{32}}, \bibinfo{pages}{165902}
	(\bibinfo{year}{2020}).
	
	\bibitem{fluer1}
	\bibinfo{author}{A.~A. Mostofi}, \bibinfo{author}{J.~R. Yates},
	\bibinfo{author}{Y.-S. Lee}, \bibinfo{author}{I.~Souza},
	\bibinfo{author}{D.~Vanderbilt}, \bibinfo{author}{N.~Marzari}.
	\newblock \bibinfo{title}{Wannier90: {A} tool for obtaining maximally-localised
		wannier functions}.
	\newblock \emph{\bibinfo{journal}{Comput. Phys. Commun.}}
	\textbf{\bibinfo{volume}{178}}, \bibinfo{pages}{685 -- 699}
	(\bibinfo{year}{2008}).
	
	\bibitem{fluer2}
	\bibinfo{author}{F.~Freimuth}, \bibinfo{author}{Y.~Mokrousov},
	\bibinfo{author}{D.~Wortmann}, \bibinfo{author}{S.~Heinze},
	\bibinfo{author}{S.~Bl{\"{u}}gel}.
	\newblock \bibinfo{title}{{Maximally localized Wannier functions within the
			{F}{L}{A}{P}{W} formalism}}.
	\newblock \emph{\bibinfo{journal}{Phys. Rev. B}} \textbf{\bibinfo{volume}{78}},
	\bibinfo{pages}{035120} (\bibinfo{year}{2008}).
	
	\bibitem{wanniertools}
	\bibinfo{author}{Q.~Wu}, \bibinfo{author}{S.~Zhang}, \bibinfo{author}{H.-F.
		Song}, \bibinfo{author}{M.~Troyer}, \bibinfo{author}{A.~A. Soluyanov}.
	\newblock \bibinfo{title}{Wanniertools: {An} open-source software package for
		novel topological materials}.
	\newblock \emph{\bibinfo{journal}{Comput. Phys. Commun.}}
	\textbf{\bibinfo{volume}{224}}, \bibinfo{pages}{405--416}
	(\bibinfo{year}{2018}).
	
	\bibitem{1TlGaTe2}
	\bibinfo{author}{A.~Nagat}, \bibinfo{author}{G.~Gamal},
	\bibinfo{author}{S.~Hussein}.
	\newblock \bibinfo{title}{Growth and characterization of single crystals of the
		ternary compound tlgate2}.
	\newblock \emph{\bibinfo{journal}{Cryst. Res. Technol.}}
	\textbf{\bibinfo{volume}{26}}, \bibinfo{pages}{19--23}
	(\bibinfo{year}{1991}).
	
	\bibitem{1InGaTe2}
	\bibinfo{author}{E.~Gojaev}, \bibinfo{author}{K.~Gyul’mamedov},
	\bibinfo{author}{A.~Ibragimova}, \bibinfo{author}{A.~Movsumov}.
	\newblock \bibinfo{title}{Conductivity piezomodulation in ingate2 single
		crystals}.
	\newblock \emph{\bibinfo{journal}{Inorg. Mater.}}
	\textbf{\bibinfo{volume}{46}}, \bibinfo{pages}{353--357}
	(\bibinfo{year}{2010}).
	
	\bibitem{1NaGaTe2}
	\bibinfo{author}{J.~Weis}, \bibinfo{author}{H.~Sch{\"a}fer},
	\bibinfo{author}{G.~Schoen}.
	\newblock \bibinfo{title}{New ternary element (i)/element (iii)-tellurides and
		selenides}.
	\newblock \emph{\bibinfo{journal}{Z. Naturforsch., B}}
	\textbf{\bibinfo{volume}{31}}, \bibinfo{pages}{1336--1340}
	(\bibinfo{year}{1976}).
	
	\bibitem{1BaPbO3}
	\bibinfo{author}{R.~Shannon}, \bibinfo{author}{P.~Bierstedt}.
	\newblock \bibinfo{title}{Single-crystal growth and electrical properties of
		bapbo3}.
	\newblock \emph{\bibinfo{journal}{J. Am. Ceram. Soc.}}
	\textbf{\bibinfo{volume}{53}}, \bibinfo{pages}{635--636}
	(\bibinfo{year}{1970}).
	
	\bibitem{1InTe}
	\bibinfo{author}{S.~Misra}, \bibinfo{author}{P.~Levinsk{\`y}},
	\bibinfo{author}{A.~Dauscher}, \bibinfo{author}{G.~Medjahdi},
	\bibinfo{author}{J.~Hejtm{\'a}nek}, \bibinfo{author}{B.~Malaman},
	\bibinfo{author}{G.~J. Snyder}, \bibinfo{author}{B.~Lenoir},
	\bibinfo{author}{C.~Candolfi}.
	\newblock \bibinfo{title}{Synthesis and physical properties of
		single-crystalline inte: towards high thermoelectric performance}.
	\newblock \emph{\bibinfo{journal}{J. Mater. Chem. C}}
	\textbf{\bibinfo{volume}{9}}, \bibinfo{pages}{5250--5260}
	(\bibinfo{year}{2021}).
	
	\bibitem{1Tl5Te2Br}
	\bibinfo{author}{D.~Babanly}, \bibinfo{author}{M.~Babanly}.
	\newblock \bibinfo{title}{Phase equilibria in the tl-tlbr-te system and
		thermodynamic properties of the compound tl5te2br}.
	\newblock \emph{\bibinfo{journal}{Russ. J. Inorg. Chem.}}
	\textbf{\bibinfo{volume}{55}}, \bibinfo{pages}{1620--1629}
	(\bibinfo{year}{2010}).
	
	\bibitem{1Tl4PbSe3}
	\bibinfo{author}{T.~Malakhovska}, \bibinfo{author}{M.~Y. Sabov},
	\bibinfo{author}{E.~Y. Peresh}, \bibinfo{author}{V.~Pavlyuk},
	\bibinfo{author}{B.~Marciniak}.
	\newblock \bibinfo{title}{Crystal structure of the tl4pbse3 ternary compound}.
	\newblock \emph{\bibinfo{journal}{Chem. Met. Alloys}} \bibinfo{pages}{15--17}
	(\bibinfo{year}{2009}).
	
	\bibitem{1Tl4SnSe3}
	\bibinfo{author}{I.~Barchij}, \bibinfo{author}{M.~Sabov},
	\bibinfo{author}{A.~El-Naggar}, \bibinfo{author}{N.~AlZayed},
	\bibinfo{author}{A.~Albassam}, \bibinfo{author}{A.~Fedorchuk},
	\bibinfo{author}{I.~Kityk}.
	\newblock \bibinfo{title}{Tl4sns3, tl4snse3 and tl4snte3 crystals as novel ir
		induced optoelectronic materials}.
	\newblock \emph{\bibinfo{journal}{J. Mater. Sci. Mater. Electron.}}
	\textbf{\bibinfo{volume}{27}}, \bibinfo{pages}{3901--3905}
	(\bibinfo{year}{2016}).
	
\end{thebibliography}
\end{document}